# Tabletop X-ray ghost video of moving objects

**Hui Zeng,[a,d,†] Ming-Fei Li,[c] Zhi-Yue Yu,[b] Bing-Zhan Shi,[a,d] Xiao-Jing Wu,[b] Jie Feng,[a,d] Jin-Guang Wang,[c] Yi-Fei Li,[c,*] Ling-An Wu,[c] Jian-Hong Shi,[b,†,*] Li-Ming Chen[a,d,*]**

[a]State Key Laboratory of Dark Matter Physics, Key Laboratory for Laser Plasmas (MoE), School of Physics and Astronomy, Shanghai Jiao Tong University, Shanghai 200240, China

[b]State Key Laboratory of Photonics and Communications, Institute for Quantum Sensing and Information Processing, School of Automation and Intelligent Sensing, Shanghai Jiao Tong University, Shanghai 200240, China

[c]Beijing National Laboratory for Condensed Matter Physics, Institute of Physics, Chinese Academy of Sciences & School of Physical Sciences, University of Chinese Academy of Sciences, Beijing 100190, China

[d]IFSA Collaborative Innovation Center, Shanghai Jiao Tong University, Shanghai 200240, China

**Abstract**. X-ray imaging is widely employed in clinical medicine, industrial inspection, and various scientific research fields. Unfortunately, most currently used X-ray two-dimensional (2D) detectors suffer from a fundamental trade-off between the number of pixels and readout time, making them unsuitable for fast moving objects imaging, as well as the readout dead time causes frame losses. X-ray ghost imaging (XGI) offers an alternative approach to image an object using only a highly sensitive single-pixel detector. However, a critical limitation of existing XGI methods is the excessive total acquisition time required, rendering it impractical for real applications. In this paper, we propose a rapid spatial modulation scheme based on random binary patterns encoded onto a fast-spinning mask. Clear X-ray visualization of moving objects is demonstrated with imaging rates up to 200 frames per second with a resolution of 225 μm. For the first time, our method has greatly improved the XGI imaging speed and paves the way for X-ray imaging application of motion objects, such as the inspection of rotating aero-engines and in vivo medical imaging.

*Address all correspondence to Yi-Fei Li, yflx@iphy.ac.cn; Jian-Hong Shi, purewater@sjtu.edu.cn; Li-Ming Chen, lmchen@sjtu.edu.cn

[†]These authors contributed equally to this work.

## 1 Introduction

X-ray imaging is an irreplaceable tool in medicine, crystallography, nondestructive inspection, and many other fields owing to its high penetration and high-resolution capabilities. Conventional direct projection imaging relies on the readout of the transmitted X-ray intensity from a two-dimensional (2D) detector. However, similar to the pixelated cameras used for other electromagnetic wavelengths, X-ray detectors face a fundamental trade-off between the number of pixels and readout time: the duty cycle is usually low, i.e., the time required for readout is much longer than the signal integration time, which usually limits applications to static objects. For example, the 1024×1024 pixels Andor X-ray camera typically has a frame rate of ~4.4 frame per second (fps)[1]. High resolution inherently leads to slower frame rates, whereas faster detectors often suffer from poor resolution or a small field of view (FOV)[2,3]. Even with advances in technology, detectors equipped with fast electronic shutters still suffer from limited frame rates, small FOV and high cost[4,5]. For example, some current so-called ultrafast X-ray cameras can achieve MHz-scale frame rates for hard X-ray imaging, but are limited to about 128 frames per burst (data acquisition burst), and slow readout to transfer the data, due to on-chip storage constraints[6]. Thus, the development of an accessible X-ray imaging detector that integrates high spatial resolution with high frame rate remains a tough challenge.

In view of the above-mentioned issues, ghost imaging (GI)[7–9], which records object information using only a single-pixel detector (SPD), can provide a promising solution. The interest in GI can be attributed to its ability to achieve imaging under very low illumination[10,11], robustness in turbulence[12], improved spatial resolution, and high signal-to-noise ratio (SNR)[13–15]. However, the technique requires successive illumination of the sample with a series of different spatial modulation patterns, then averaging over the corresponding correlation between the transmitted or reflected light from the object with each modulation pattern. This lengthy exposure process and data accumulation and computation time are obvious disadvantages, but recently several high-speed modulation schemes based on spinning masks have been proposed and demonstrated for single-pixel imaging of moving objects at visible wavelengths[16–20].

In general, GI is applicable to any wavelength and has already been realized across the entire spectrum, from microwaves[21] to terahertz[22,23], hard X-rays[10,13,24,25], and even with atoms[26,27], electrons[28,29], and neutrons[30–33]. Compared to visible light and near-infrared, XGI research started relatively late. The major difficulty lies in the lack of suitable optics for focusing and intensity modulation. To address this problem, two experimental approaches have been proposed: one is to utilize a crystal as a beam splitter, which generally requires a very high-flux X-ray source[25,34–36]; the other is to use modulation masks made of highly absorptive materials to generate a series of intensity-modulated two-dimensional patterns for illuminating the object, such as sandpaper[10], copper[13], gold[13], copy paper[34], glass[36], or metal powder[35]. Various forms of XGI, such as X-ray computed tomography[35], phase-contrast imaging[37,38], fluoroscopy[39–41] and absorption spectra[42] have already been demonstrated. Meanwhile, some studies employing the parallel ghost imaging scheme to expand the imaging FOV and improve resolution have also achieved success[30,43–46]. However, the modulation of discrete patterns in the aforementioned schemes results in a relatively long total time required to expose all measurement patterns one by one, which can be as long as several hours. The first XGI experiment for moving objects was demonstrated by using sandpaper to introduce pre-recorded intensity fluctuations[47]. However, the procedure and results described in that study still required a very long sampling time and were only applicable to a limited scope, i.e., imaging periodically moving objects. In our previous work[48], we implemented a fast XGI scheme by reshaping a cyclic S-matrix into an aggregate pattern and performing continuous translational scanning, which reduced the imaging time to the order of seconds. However, real time dynamic imaging is still challenging due to limited modulation speed.

In this work, we present a new method for 2D high-resolution XGI of moving objects by combining spinning disk modulation with compressed-sensing (CS)[49], demonstrating pattern modulation rates of up to 204.8 kHz. Our scheme is based on a specially designed rotating annular binary mask with cyclic random patterns. During one revolution, the mask illuminates the object with a series of modulation patterns that enable reconstruction of one image. The imaging frame rate is therefore determined by and equal to the revolution speed. In our work, the mask is driven by a high-speed motor to spin at a constant speed of up to 200 revolutions per second (rps), achieving an imaging frame rate of 200 fps for images of 1024 pixels. Through this scheme, the letters on a rotating target have been successfully imaged and recorded.

## 2  Materials and Methods

*2.1 Experimental Setup*

The experimental setup is shown in Fig. 1. An X-ray tube (Incoatec Source IusCu) was operated at 35 kV and 800 µA as the light source. A tungsten aperture with a size of 5.6×5.6 mm² was positioned downstream of the tube to ensure that only a single pattern of the coding mask was illuminated. Given the slight divergence of the beam, the distance between the aperture and the mask was set to 15 cm, projecting an approximately 7.2×7.2 mm² square beam onto a subarea of the modulation pattern. The mask was driven by a motor to spin continuously at a constant speed of 200 rps; the maximum jitter was 0.04% per cycle. The object, a copper disk on which various letters were stencilled in an annular arrangement, was placed as close as possible behind the mask (at a distance of ~4 cm) to minimize blurring of the encoded pattern caused by the uncollimated beam. As both the mask and object were rotated, the letter was successively illuminated by different segments of the mask pattern. The total intensity transmitted through the object was then measured in real time using an economical X-ray SPD (North Night Vision Technology, Nanjing, Model M3021), which consists of a sodium iodide scintillator, photomultiplier tube, high-voltage module, and preamplifier. This detector has an effective detection diameter of 25 mm, a detection efficiency of 82%, and a response time of approximately 2 μs for X-rays in the 5–100 keV energy range. Its operating voltage was set to 650 V. To synchronize the coding patterns with the transmitted signals, a light emitting diode (LED) and a photodiode (PD) were aligned on opposite sides of the mask, which had a notch at its edge so that a light pulse from the LED could be received by the PD once per rotation. Both the transmitted signals from the SPD and the PD signals were synchronously recorded by a data acquisition card (TiePie Handyscope HS6 oscilloscope), which recorded the fluctuating voltage-time signals and transmitted them to a computer. The sampling frequency was set to 1 MHz to ensure that all photons reaching the detector could be collected. After correlating the encoded patterns with their corresponding voltage signals, the object image was reconstructed. An X-ray sCMOS camera (Photonic Science Large Area VHR22M_125, pixel size 25 μm, 250 μm thick CsI scintillator) was also used to take conventional photos of the object disk.

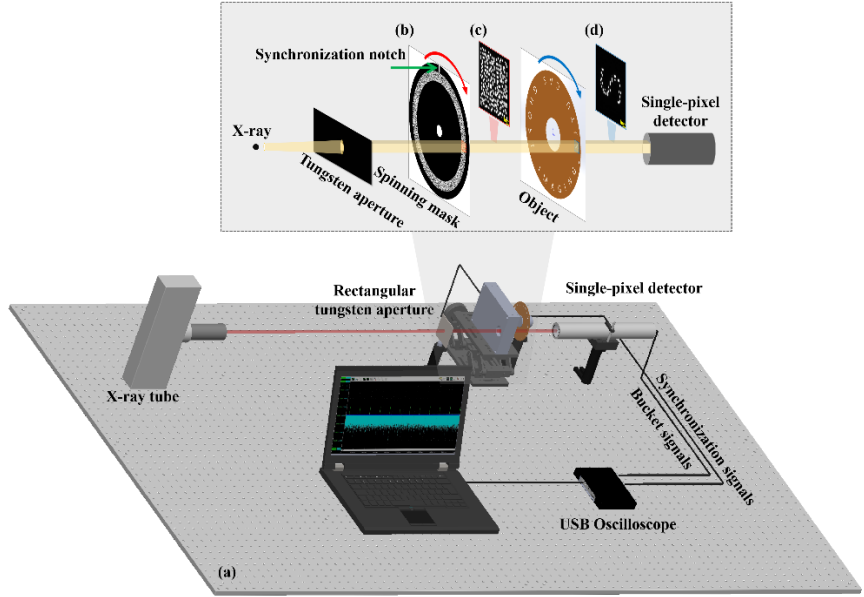

**Fig. 1** Diagram of the experimental setup. (a) Overall setup. (b) The spinning coding mask. The red square indicates the imaging area, while the green arrow indicates the synchronization notch. (c) and (d) Intensity distributions of the beam just behind the mask and behind the object recorded by an X-ray sCMOS camera, respectively.

*2.2 Spinning Coding Mask*

The design of the modulation mask is shown in Fig. 2(a). As shown in Fig. 2(b), a rectangular random binary code was condensed into an aggregate pattern consisting of 25% transmitting (1) and 75% blocking (0) pixels of size 32×1024, then stencilled in a circular pattern on a 500 μm thick brass disk. We note that a new pattern is illuminated each time the scanning area moves one pixel forward, see Figs. 2(b) and (c) where three of them are outlined in the red, green, and yellow dotted boxes. The random matrix patterns have the advantage of avoiding decimation when applying CS algorithms[50,51]. The imaging spatial resolution is determined by the size of the pattern element, which is 225 μm in our mask. The 0.5 mm thickness of the disk enables significant absorption of X-ray photons of energies up to 35 keV. In practice, the intensity modulation of the mask cannot be strictly 0 or 1, as it depends on the absorption of the material, so it is important to measure the actual modulation depth which determines the contrast-to-noise ratio (CNR) of the final image[52,53]. This was performed by measuring the total intensity over a FOV of 7.2×7.2 mm² with the SPD with the beam completely unblocked, and then measuring the intensity after the beam passed through an uncoded portion of the brass mask. From the results shown in Fig. 2(d), we calculated the modulation depth to be approximately 98.93%, which is much higher than that in Ref. 48.

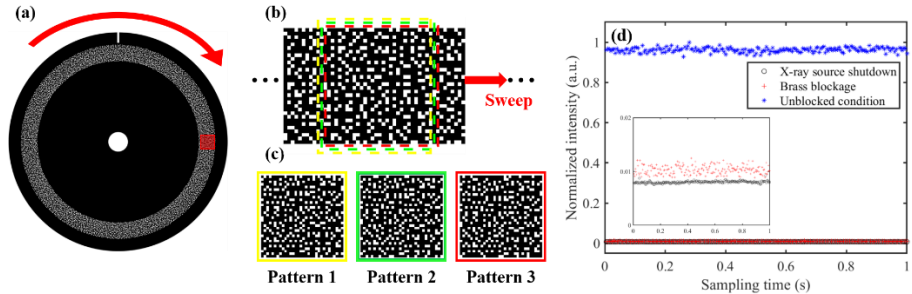

**Fig. 2** Modulation mask. (a) Annular modulation mask. The red arrow illustrates the spin direction, and the red box the segment illuminated after the aperture. (b) Schematic of modulation pattern sweeping for a frame size of 32×32 pixels, with colored boxes representing the successively sampled patterns. Red arrow indicates scan direction. (c) Three representative successive encoding patterns. (d) The normalized intensity measured by the SPD with the X-ray beam completely unblocked (top blue stars), and with the beam passing through an uncoded portion of the brass mask (bottom red "+"). The detector's sampling time and frequency are 1 s and 1 MHz, respectively. Each integration time is 5 ms.

## 3  Results and Discussion

### 3.1  Imaging of Static Objects

We first used our mask to perform XGI on the object in a stationary state; the results of which are shown in Fig. 3, where the object is depicted in Fig. 3(a); the stencilled letters are "GHOST IMAGING SJTU CAS". Figures 3(b1)–(b4) present the projection images of a portion of the letters taken by the sCMOS camera, and the images reconstructed by XGI are shown in the row below in Figs. 3(c1)–(c4). The white regions correspond to the full transmission of X-rays through the cut-out parts of the object. In the XGI the coding mask was rotated at a speed of 200 Hz. To quantitatively evaluate the quality of the reconstructed XGI images, the CNR (the definition is specified in Supplementary Note 1) was calculated[53], and is shown in the upper left-hand corner of Figs. 3(c1)–(c4). The image reconstruction methods are given in Appendix A. Both the projection and XGI images were acquired under the same X-ray source parameters and total exposure time (i.e., 1 s). The normalized intensities of the cross-sections marked by the green vertical lines and red horizontal lines in Figs. 3(c1)–(c4) are plotted in Figs. 3(d1)–(d4). By measuring the half-width of the slope of the cross-section near the edges of the image[45], the spatial resolutions in the horizontal and vertical directions are estimated to be approximately 225 μm, as shown in Fig. 3(e), which agrees well with the 225 μm width of the modulation pixels. Additionally, the corresponding gray-scale imaging results are provided in Supplementary Note 2.

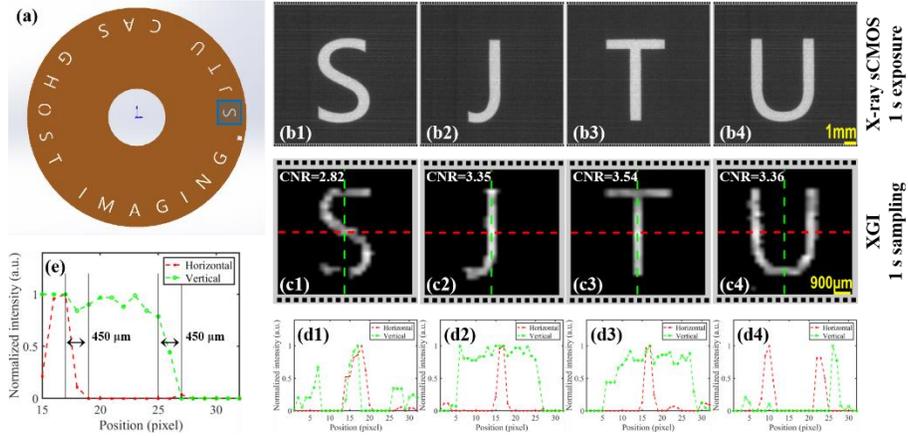

**Fig. 3** Experimental results for the static object. (a) Photo of the object disk; the blue box indicates the region of interest. (b) Projection X-ray images of a portion of the static object taken by the sCMOS camera with 1 s exposure time. (c) Reconstructed XGI images for the same sampling time of 1 s; their corresponding CNR values are shown at the top left-hand corner. (d) Normalized intensity profiles along the red and green lines of Fig. 3(c). (e) Detailed magnified view of a segment from Fig. 3(d2).

Different total sampling times correspond to different total numbers of photons. To investigate the influence of photon counts on the CNR, we varied the sampling time for each reconstruction from 50 ms to 20 s. The reconstructed images are presented in Figs. 4(a)–(f), with their corresponding CNR curves shown in Figs. 4(g) and 4(h). Analysis of the images row by row reveals that as the sampling time increases from 50 ms to 1 s, the CNR increases significantly due to greater abundance of the signal photons. However, from 1 s to 20 s, little variation is observed, and the CNR values tends to saturate.

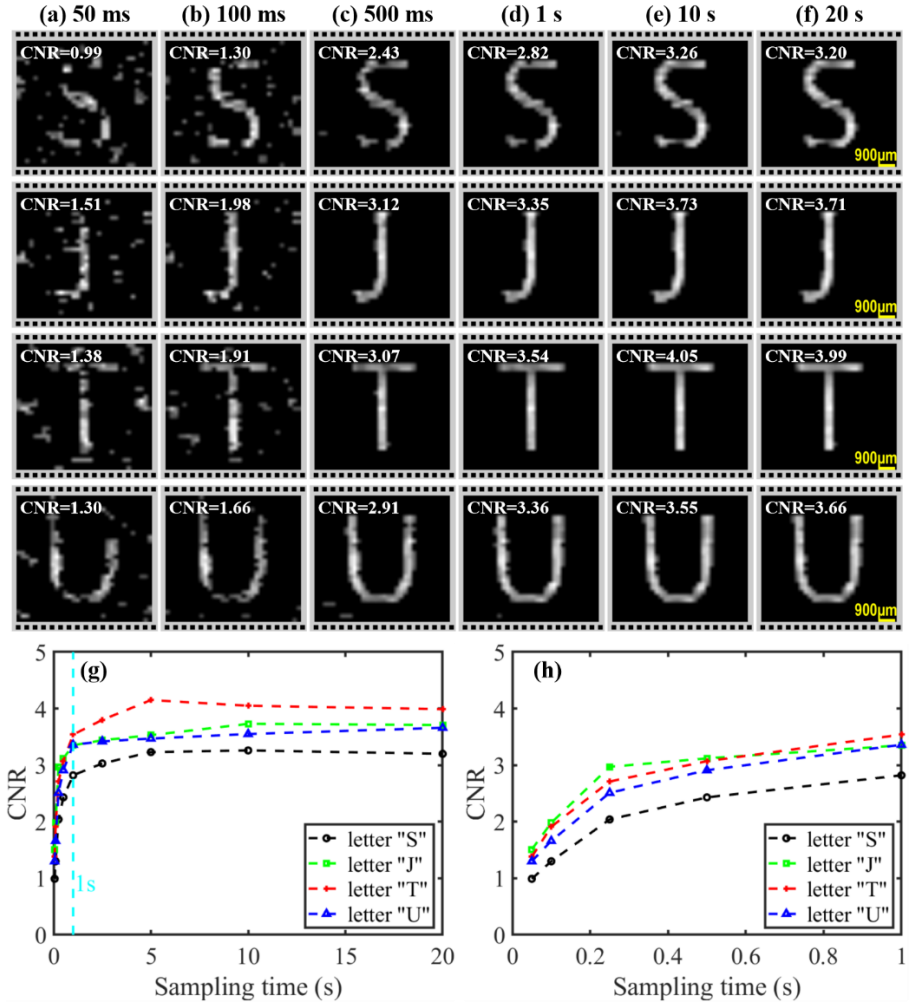

**Fig. 4** Comparison of image reconstruction for different sampling times. (a)–(f) The images retrieved by XGI for different sampling times (50 ms to 20 s). (g) The CNR curves corresponding to Figs. (a)–(f). (h) Detailed magnified view of the 50 ms to 1 s range from Fig. 4(g).

## 3.2  Dynamic Videos of Rotating Objects

The stencilled object disk was driven by a motor to rotate at a series of constant speeds of 1/256, 1/64, 1/16 and 1/4 rps; since the radial distance of the letters was 4.75 cm, this corresponded roughly to the letters moving vertically downwards at velocities of 0.12, 0.48, 1.92 and 7.68 cm/s, respectively. Imaging with our scheme was performed using 1024 rectangular illumination patterns of size 32×32 pixels. The fully captured videos at different object velocities are available in Videos S1-S7. The modulation mask was rotated at 200 rps, so the videos were all reconstructed at an imaging speed of 200 fps with no dead time; selected frames from each video are presented in Figs. 5(a)–(d), and full details of the captured videos are provided in Note 3 in the Supplementary Material. A key limitation of our scheme is the effective X-ray photon flux; the limited photon flux of the tabletop X-ray source (approximately $10^5$ photons/mm²·s at the object plane) results in smaller modulation fluctuations in the X-ray intensity over short sampling intervals, which are difficult to detect. Additionally, for short sampling durations, noise from random cosmic rays, background photons and readout circuitry degrades image quality. Stacking multiple modulation cycles remains necessary for modulation under low flux conditions to improve the imaging SNR. Meanwhile, a motion compensation algorithm (see Appendix A for more details) was developed to correct misalignment between the modulation matrix and bucket signals, thereby eliminating motion blur effects. The benefit of motion compensation in terms of motion blur and image quality is demonstrated in Supplementary Note 4.

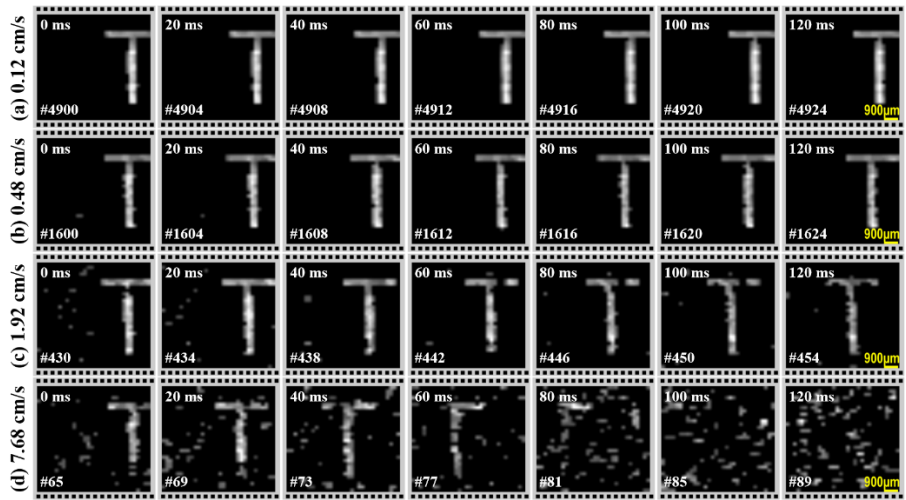

**Fig. 5** Snapshots from the XGI videos of the moving letter T. (a)–(d) Object velocities of 0.12, 0.48, 1.92 and 7.68 cm/s; the serial number of the video frame is shown at the bottom-left corner of each image.

A comparison of images across the columns in Fig. 5 reveals that the overall image quality degrades progressively as the object velocity increases from 0.12 cm/s to 7.68 cm/s. This degradation is due to the reduced residence time of the object within the FOV; as shown in Fig. 5(d), an object moving at 7.68 cm/s has a residence time of only about 90 ms when traversing the 7.2 mm FOV, which in turn reduces the number of possible stacking cycles. Higher-quality imaging of faster-moving objects can be achieved by increasing the source intensity and/or improving the detector response speed. Furthermore, once the source flux and sampling frequency are optimized, the imaging frame rate can be increased by increasing the mask rotation speed, provided that a sufficient number of photons are detected at the detector. Apart from the source intensity and detection system, another factor that can be limiting imaging quality is instability in the mask rotation speed or off-axis deviation of the mask relative to the motor axis. Reducing the sampling rate (defined as the ratio of the number of sampling patterns employed to the total number of patterns) is another method to further increase the imaging rate. As described in Note 5 in the Supplementary Material, random matrices offer the advantage of enabling CS to reconstruct high-quality images. Even with signal truncation, the object remained clearly visible when only 25% or even 12.5% of the measurement data was used. Correspondingly, the imaging rate can be increased by at least a factor of 8 (i.e., up to 1.6 kfps) by using only 1/8 of the sampling matrices and reconstructing the image using CS algorithms.

*3.3  Comparison with Array Camera Imaging*

To evaluate the unique capability of dead-time-free video recording of our system for imaging moving objects, we conducted experiments to compare its performance with that of traditional projection-based cameras. For a fair comparison, 8×8 on-chip binning was applied to the X-ray sCMOS camera for a resolution of 200 μm—comparable to the 225 μm resolution of our XGI system. Correspondingly, the imaging frame rate of the camera was increased from 1 fps to 8 fps. Figure 6(a) displays the letters GHOST extracted from the XGI video at a frame rate of 200 fps, while Figs. 6(b) and 6(c) show several consecutive frames of the letters captured by the sCMOS camera with a 5 ms exposure. As shown in Fig. 6(b), the array camera can capture high-clarity images but suffer from frame losses due to the required readout time after each exposure. This frame loss issue caused by dead time is widespread in most commonly used array cameras, leading to incompleteness and possible loss of critical information. This is illustrated in Fig.6(c) where the object speed is increased to 15.36 cm/s. Another comparison with a commonly used 1024×1024 pixels (13 μm pixel size) Andor iKon-M SO X-ray CCD is provided in Figs. 6(d) and 6(e); the frame rate increased from 4.4 fps to 25 fps with 16×16 binning, corresponding to an imaging resolution of 208 μm. From these comparisons, it can be concluded that as the imaging rate increases, the captured information can be richer, but frame loss remains inevitable. Additional comparisons under more object velocities can be found in Note 6 in the Supplementary Material. In contrast, our XGI system performs continuous measurements, thus enabling complete recording of all the moving letters, despite the relatively poor imaging quality, which is mainly due to the low radiation flux that we used. On another note, it should be point out that the image processing time of current array cameras is limited by the number of pixels, so binning enhances the imaging speed and also improves the SNR, but resolution is lowered.

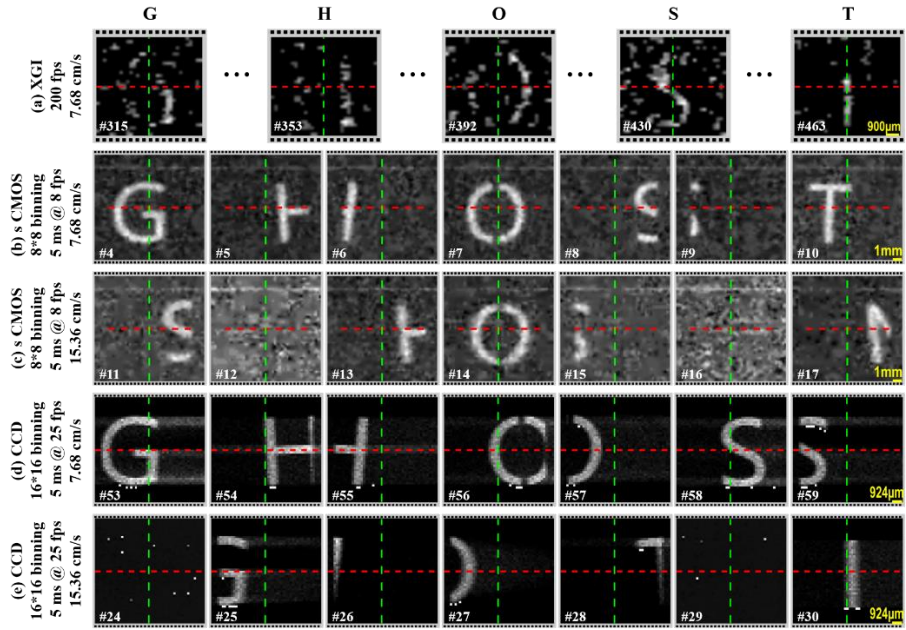

**Fig. 6** Frame loss phenomenon due to the unavoidable dead time of the array camera. (a) Selected partial frames of a video reconstructed by our XGI camera for the object velocity of 7.68 cm/s. (b) and (c) Several consecutive frames captured by the sCMOS camera for velocities of 7.68 cm/s and 15.36 cm/s, respectively. (d) and (e) Andor CCD results corresponding to Fig. 6(b) and (c), respectively. The red and green dashed lines indicate the center of the FOV.

## 4  Conclusions

In conclusion, we have developed an XGI system capable of high-resolution imaging of fast-moving objects. Rapid spatial modulation rates of up to 204.8 kilopixels/s is achieved by rotating an annular brass mask encoded with a random binary pattern, thereby reducing the imaging time to several milliseconds. A modulation depth close to 1 enabled us to obtain images with significantly higher CNR with a simple tabletop X-ray source. Video-rate imaging was achieved at rates up to 200 fps, allowing us to capture objects moving at speeds of up to 7.68 cm/s. The method and results described herein open new possibilities for the study of dynamic processes.

In principle, the imaging speed of our approach is mainly limited by the rotation speed of the mask, as well as the bandwidth (i.e., sampling frequency) of the detection system. Currently constrained by the X-ray flux used in our experiments, the imaging rate could be further improved by increasing the radiation source intensity, but preferably by enhancing detector sensitivity and response speed. Additionally, for specific applications where spatial resolution is crucial this could be enhanced by reducing the modulation pattern pixel size. With micro/nano fabrication and high-precision lithography techniques, achieving both nanometer-scale resolution and large pixel counts while maintaining the high modulation depth should be feasible in the near future; selecting modulation materials with high X-ray absorption properties is another viable approach. Optimizing these parameters would enable further improvements.

Notably, our proposed method is not limited to any wavelength but would be particularly suitable for situations where array cameras are cumbersome, extremely expensive and difficult to maintain. In particular, in contrast to other available X-ray imaging techniques for moving objects, our method can be implemented with a conventional tabletop X-ray source. The setup is simple, relatively inexpensive, and easy to operate. Hence, our method holds great potential for applications in medical imaging where physical movement exists (e.g., non-invasive cardiac imaging) and in non-destructive imaging of moving mechanical components, such as high-speed rotating aircraft engines and turbine blades.

**Appendix A: Reconstruction Methods**

*A.1 Image Algorithm*

Random sampling inherently exhibits efficient, universal, and robust characteristics in signal acquisition[50,51]. In our work, we employ random binary codes as the sampling matrix, which enables the reconstruction of high-quality images via CS algorithm[49]. Here we adopt the augmented Lagrangian and alternating direction based total variation algorithm TVAL3[54] to reconstruct the images:

$$\min_{\mathbf{x}} \sum_i \|D_i \mathbf{x}\|_2 + \frac{\mu}{2} \|A\mathbf{x} - \boldsymbol{b}\|_2^2, \quad s.t. \quad \mathbf{x} \geq 0,$$

where $\|\mathbf{x}\|_2$ is the $l_2$ norm of vector $\mathbf{x}$, $\mathbf{x} = [x_1, \dots, x_n]^T$ denotes the estimated object vector, $\mathbf{b} = [b_1, \dots, b_m]^T$ ($m \leq n$) represents the bucket intensity integrated for a given modulation times, $\mathbf{A}$ is the modulation matrix of size $m{\times}n$, $\mu>0$ is the penalty parameter of the quadratic fidelity term, $D_i\mathbf{x}$ is the $i^{th}$ component of the discreet gradient of $\mathbf{x}$, and $\sum\|D_i\mathbf{x}\|_2$ denotes the total variation. The fidelity term $\|A\mathbf{x}-\mathbf{b}\|_2^2$ minimizes the difference between the estimation and measurements, while the total variation removes the image noise and preserves edge information at the same time.

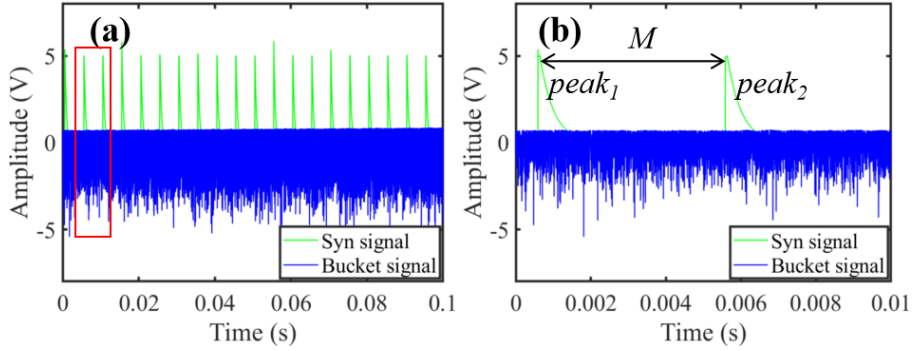

**Fig. 7** Typical bucket (blue) and synchronization (green spikes) signals. (a) A data train containing 20 modulation cycles. (b) Magnified view of the red box in (a), showing 2 synchronization pulses.

In our experiment, the rotation speed of the mask was fixed at 200 Hz; and the detector sampling speed at 1 MHz. Fig. 7 shows a typical example of the bucket and synchronization signals recorded by our acquisition system. The data between every two adjacent synchronization signal spikes corresponds to one set of continuously modulated signal during one complete rotation of the mask, as illustrated in Fig. 7(b). To efficiently transform this data into an image with 1024 pixels, we adopt interpolation operations and then evenly distribute the interpolated data among the bucket signals **b** which correspond to the 1024 sequential modulation matrices. To enhance the SNR of the reconstructed images, a data multiplexing method was introduced into our reconstruction algorithm. We repeated the above processing steps for the subsequent periodic data and summed the corresponding bucket signals within each period to obtain the final bucket values.

*A.2 Motion Compensation*

The relative motion between the mask and the object induces a misalignment between the modulation matrix **A** and the bucket signals **b**, thereby introducing a motion blur effect in the data multiplexing-based reconstructed image, so it is important to compensate for this. In our experiment, when the object being imaged is in a stationary state, the data acquisition card records $M$ points during one revolution of the mask. For a moving object, the object and mask rotate in the same direction, with the object velocity ($v_{obj}$) being $1/k$ of the mask velocity ($v_{mask}$). This means that the measurement cycle becomes longer. When the mask completes one revolution, the corresponding data collected by the acquisition card increases by an offset of $1/k$ (i.e., $M + (M/k)$) relative to the stationary state, which leads to motion blur in image reconstruction when multiple cycles are stacked to improve SNR. Therefore, this offset must be compensated for to realign the 1024 modulation patterns with each object letter bucket signals.

In image reconstruction framework, a motion compensation algorithm was integrated to realign the bucket signals **b**. As illustrated in Fig. 7, the synchronization signal pulses (green curve) provide critical temporal markers (e.g., $peak_1$, $peak_2$, $peak_3$, …) defining the nominal modulation cycles. The compensation algorithm operates as follows:

(1) With the mask speed fixed at 200 Hz, i.e., $v_{mask} = 200$ rps. And the object velocity $v_{obj}$, e.g., 0.5 rps, the $k$ is equal to:

$$k = \frac{v_{mask}}{v_{Obj}} = 400,$$

(2) The nominal modulation cycle $M$ of a full modulation (e.g., $M = peak_2 - peak_1$) is extracted directly from the synchronization signal.

(3) Due to relative motion, the actual data segment corresponding to the object modulation within each nominal cycle requires an offset for retrieval. The data retrieval for each cycle is shifted backward by an amount $M/k$ relative to the nominal peak positions. Consequently, the boundaries for the actual modulation cycle are defined as:

**Cycle 1:** $(peak_1) \sim (peak_2 + (M/k))$;

**Cycle 2:** $(peak_2 + (M/k)) \sim (peak_3 + 2 \times (M/k))$;

**Cycle 3:** $(peak_3 + 2 \times (M/k)) \sim (peak_4 + 3 \times (M/k))$;

……

**Cycle N:** $(peak_N + (N-1) \times (M/k)) \sim (peak_{N+1} + N \times (M/k))$;

and so on.

(4) The final bucket signal **b** for each frame reconstruction is formed by bit-wise summation of **Cycle 1** with subsequent cycles (e.g., **Cycle 2** and **Cycle 3**, …, **Cycle N**). Here, *N* is determined by the residence time of the object within the FOV.

This compensation not only eliminates motion blur but also improves the imaging SNR.

*Disclosures*

The authors declare that there are no financial interests, commercial affiliations, or other potential conflicts of interest that could have influenced the objectivity of this research or the writing of this paper.

*Code, Data, and Materials Availability*

Data underlying the results presented in this paper are not publicly available at this time but may be obtained from the authors upon reasonable request.

*Acknowledgments*

We acknowledge funding for this work from the National Natural Science Foundation of China (12335016, W2412039, 11991073, 12305272) and the Strategic Priority Research Program of the Chinese Academy of Sciences (XDA25030400, XDA25010100, and XDA25010500). The experiment was carried out at the Synergetic Extreme Condition User Facility, Beijing (SECUF, http://cstr.cn/31123.02.SECUF.D3).

**Hui Zeng** is currently a Ph.D candidate; he started his graduate research in 2022 in the field of X-ray ghost imaging at the Key Laboratory for Laser Plasmas (MoE), Shanghai Jiao Tong University.

**Ming-Fei Li** received his PhD from the Institute of Physics, Chinese Academy of Sciences (CAS) in 2014, and is currently a professor there in the Center for Applied Physics. His research is focused on ghost imaging/single-pixel imaging across all wavelengths.

**Yi-Fei Li** received his PhD from the Institute of Physics (CAS) in 2017. His research interests include novel laser-driven x/γ radiation sources and its application, as well as laser driven electron acceleration.

**Jian-Hong Shi** received his PhD in optics in 2003 from the Department of Physics of Shanghai Jiao Tong University, where he is currently an associate professor. His research focuses on quantum correlation imaging and single-photon imaging technology.

**Li-Ming Chen** is a professor at Shanghai Jiao Tong University, before which he was a professor at the Institute of Physics, CAS. He graduated from Fudan University in 1990, and received his Ph.D. degree in Plasma Physics from the Institute of Physics, CAS in 2000. His current research interest is focused on laser driven electron acceleration, ultrashort X-ray generation and its applications.

**Ling-An Wu** obtained her PhD from the University of Texas at Austin in 1987, and is a professor emerita in the Institute of Physics (CAS). Her current interest is focused on ghost imaging in the X-ray region.

Biographies and photographs for the other authors are not available.